\documentclass[twocolumn,aps,prl,showpacs]{revtex4}
\usepackage{graphicx}
\usepackage{epsfig}

\begin{document}

\title{Supersolid phases in the one dimensional extended soft core
  Bosonic Hubbard model}

\author{G.G. Batrouni$^1$, F. H\'ebert$^1$, R.T. Scalettar$^2$} 
\affiliation{$^1$Institut Non-Lin\'eaire de Nice,  UMR 6618 CNRS,
  Universit\'e de   Nice--Sophia Antipolis, 1361 route des Lucioles,
  06560 Valbonne, France} 
\affiliation{$^2$Physics Department, University of California, Davis, CA
  95616, USA} 

\begin{abstract}
We present results of Quantum Monte Carlo simulations for the soft
core extended bosonic Hubbard model in one dimension exhibiting the
presence of supersolid phases similar to those recently found in two
dimensions. We find that in one and two dimensions, the
insulator-supersolid transition has dynamic critical exponent $z=2$
whereas the first order insulator-superfluid transition in two
dimensions is replaced by a continuous transition with $z=1$ in one
dimension. We present evidence that this transition is in the
Kosterlitz-Thouless universality class and discuss the mechanism
behind this difference. The simultaneous presence of two types of
quasi long range order results in two soliton-like dips in the
excitation spectrum.
\end{abstract}


\maketitle

Experimental and theoretical work on supersolidity (SS), a
thermodynamically stable phase in which diagonal (density) and
off-diagonal (superfluid) order coexist~\cite{bunch}, has seen an
enormous resurgence over the past several years.  In the case of solid
Helium, the observation of an anomalous moment of
inertia~\cite{kimchan} has been vigorously re-investigated by several
groups both experimentally~\cite{expHe} and theoretically~\cite{thHe}.
In parallel, an entirely new arena for the investigation of supersolid
phenomena has arisen in the context of cold atoms in optical lattices,
whose description by the bosonic Hubbard Hamiltonian is well
established~\cite{jaksch}.  It is known that this model does not
exhibit supersolid phases due to the absence of longer range
interactions.  However, the recent realization~\cite{pfau} of
Bose-Einstein condensate of $^{52}Cr$ atoms, which have an
exceptionally large magnetic dipole moment and therefore long-range
interactions, has opened wide the possibility of experimantal
observation of SS phases on confined optical
lattices~\cite{optlatice}.

Such strongly dipolar ultra-cold atoms, when placed on optical
lattices, are governed by the {\it extended} Bosonic Hubbard
Hamiltonian which has been shown, via extensive Quantum Monte Carlo
(QMC) simulations, to exhibit SS phases in two dimensions. The model
we consider is,
\begin{eqnarray}
\nonumber
H&=&-t\sum_{<ij>} \left(a^{\dagger}_i a_{j} + a^{\dagger}_{j} 
 a_i\right) + \frac{U}{2}\sum_i n_i(n_i-1) \\
& & + V\sum_{<ij>} n_i n_j.
\label{hubham}
\end{eqnarray}
The hopping parameter, $t$, sets the energy scale, $n_i=a^\dagger_i
a_i$ and $[a_i,a^\dagger_j]=\delta_{ij}$ are bosonic creation and
destruction operators. The repulsive contact (near neighbor)
interaction is $U$ ($V$).
Recently, it was shown~\cite{pinaki} in $2d$ that the model described
by Eq.~(\ref{hubham}) has a SS phase if $U$ and $V$ are such that when
the charge density wave (CDW) phase is doped {\it above} half filling,
the added bosons go on already occupied sites (roughly $U<4V$). Doping
{\it below} half filling provokes phase separation and a first order
transition to a SF phase. If multiple occupancy is suppressed, there
is phase separation when the system is doped above half filling as in
the hard core case\cite{ggb1}. This raises the possibility that an
analogous phase exists for the one dimensional form of
Eq.~(\ref{hubham}) in which case this would be the first example of a
SS in a one dimensional system.

In this letter we re-examine the one dimensional soft core bosonic
Hubbard model~\cite{white,ggb3}, Eq.~(\ref{hubham}) under the same
conditions.  We conclude that a ground state SS phase does exist with
similarities to, and important differences from, that in $2d$.
Calculations of the excitation spectra and Landau critical velocity
demonstrate that the SS phase is stable, and allow us to infer the
dynamic critical exponents.  We use the World Line algorithm in our
QMC simulations.

\begin{figure}
\psfig{file=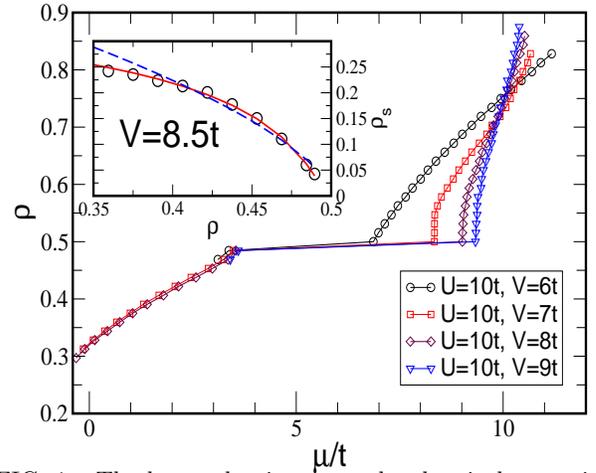,height=3.5in,width=3in,angle=-90}
\vskip-0.3in
\caption{ The boson density, $\rho$, vs the chemical potential, $\mu$,
for $L=64$, $\beta=20$ with $200$ time slices. Inset: $\rho_s$ vs
$\rho$ as $\rho=1/2$ is approached from below. See text near
Eq.~(\ref{essential}).}
\label{rhomu}
\end{figure}

We begin by demonstrating that for $V$ sufficiently large, the system
has gapped, incompressible CDW insulating behavior at half-filling,
$\rho=1/2$.  Figure~\ref{rhomu} shows $\rho$ vs $\mu$ for $U=10t$ and
$V=6t,\,7t,\,8t,\,9t$ using the WL algorithm.  The plateau at $\rho
=1/2$ clearly indicates a non-zero gap for all four cases.  However,
the behavior of the system depends on whether it is doped below or
above $\rho=1/2$, and on the value of $V$.  Starting with the CDW and
{\it removing} a particle creates a ``defect'' with three adjacent
unoccupied sites which then splits into two mobile two-hole bound
states (solitons) which destroy the long range CDW order in favor of
quasi-long range order with $\rho_s\neq 0$. Consequently, all the
curves coincide for $\rho<1/2$ in Fig~\ref{rhomu}. In
contradistinction, when the $2d$ CDW phase is doped below half
filling, phase separation takes place leading to a first order phase
transition.  Upon {\it adding} a particle to the CDW for $V\leq 6t$,
the extra particle goes to an unoccupied site since the cost of
multiple occupancy exceeds that of near neighbors. Thus, a defect of
three adjacent occupied sites is created which then splits into two
mobile two-particle bound states (solitons) which are the
particle-hole symmetric analogs of the $\rho<1/2$ case. This
particle-hole symmetry, seen for $V=6t$ in Fig.~\ref{rhomu}, is of
course approximate for finite $U$. On the other hand, upon adding a
particle to the CDW when $V>6t$, the cost of near neighbors exceeds
that of multiple occupancy, and the extra particle goes to an already
occupied site (which cannot happen in the hardcore case). This
breakdown of particle-hole symmetry is seen for $V=7t, 8t, 9t$ in
Fig.~\ref{rhomu}. For $\rho>1/2$ the curve first rises very sharply
hinting at a diverging compressibility,
$\kappa=\partial\rho/\partial\mu$ which is not the case for
$\rho<1/2$.  We shall return to this point below.

\begin{figure}
\psfig{file=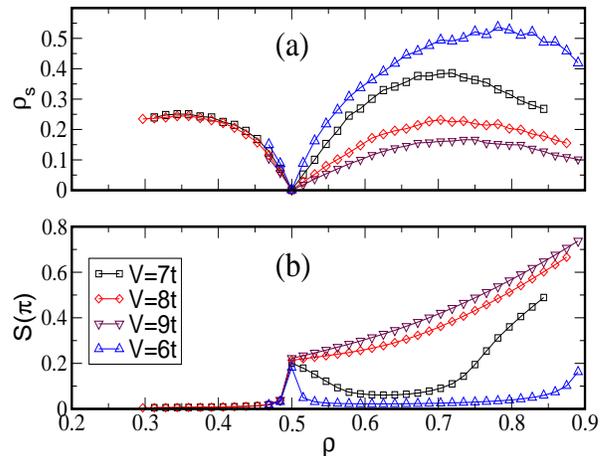,height=3.5in,width=3in,angle=-90}
\vskip-0.3in
\caption{(a) The superfluid density, $\rho_s$, and (b) the structure
  factor, $S(k)$, as functions of the number density $\rho$ for four
  values of $V$. $L=64$ sites, $\beta=20$ and $U=10t$.}
\label{rhosSpi}
\end{figure}

In order to characterise the phases, we calculate the SF
density, $\rho_s$~\cite{pollock}, and the structure factor, $S(k)$,
\begin{eqnarray}
\rho_s &=& \frac{\langle W^2\rangle}{2t\beta L^{d-2}},\\
\label{rhos}
S(k) &=& \frac{1}{L^2} \sum_{x,x^\prime} {\rm e}^{ik(x-x^\prime)}\langle
n(x)n(x^\prime) \rangle,
\label{sk}
\end{eqnarray}
where $W$ is the winding number, $L$ the length of the system, $d$ the
dimension and $\langle n(x)n(x^\prime) \rangle$ is the density-density
correlation function.  Figure \ref{rhosSpi}(a) shows $\rho_s$ versus
$\rho$ for $U=10t$ and $V=6t, \,7t,\,8t,\,9t$ while \ref{rhosSpi}(b)
shows $S(k=\pi)$ for the same values. For $\rho < 0.5$, $S(\pi)$ is
essentially zero, {\it i.e.}~there is no CDW, while at the same time
$\rho_s\neq 0$.  This is a SF phase. This behavior is {\it
qualitatively different} from that in two
dimensions~\cite{ggb1,pinaki} where the system undergoes phase
separation and a first order transition. Such phase separation cannot
happen in one dimension, and indeed Fig.~\ref{rhomu} does not exhibit
any negative compressibility regions~\cite{ggb1}. Instead, solitonic
excitations are produced~\cite{ggb3}. 

For $\rho>0.5$ and $V>6t$, $S(\pi)$ takes non-zero values indicating
simultaneous co-existence of SF and long range CDW order, in other
words a SS in one dimension. That this SS does not undergo phase
separation is again clear from Fig.~\ref{rhomu} where~\cite{ggb1} the
compressibility, $\kappa$ is never negative. For $V=7t$, finite size
scaling shows that in the thermodynamic limit, $\rho_s$ maintains a
non-zero value whereas the behavior of $S(\pi)$ depends on $\rho$ as
$L\rightarrow \infty$.  In other words, as this system is doped above
half filling, it first enters a SS phase, then a SF phase and finally
back into a SS. The phase diagram in two dimensions (Fig. 1 in
Ref.~\cite{pinaki}) does not exhibit such re-entrant behavior. On the
other hand, for $V=8t, 9t$, the system remains in the SS phase for all
$\rho > 0.5$, at least up to the highest density we studied, $\rho=
1.1$, except at $\rho=1$ where it is a gapped CDW insulator with
alternating empty and doubly occupied sites.

\begin{figure}
\psfig{file=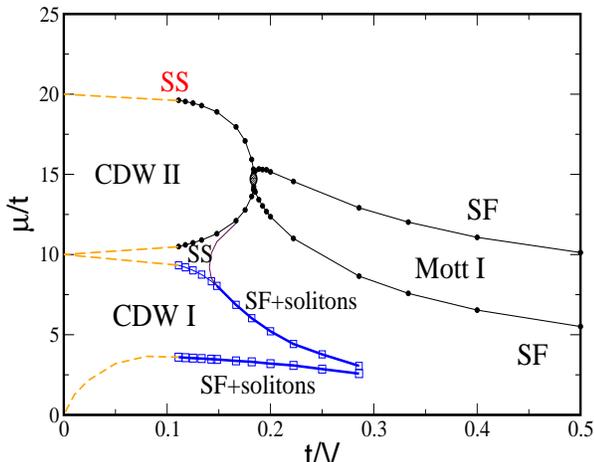,height=3.5in,width=3in,angle=-90}
\vskip-0.3in
\caption{The phase diagram for $U=10t$. CDW-I is alternating empty and
singly occupied sites, CDW-II is alternating empty and doubly occupied
sites, Mott-I is the first Mott lobe with $\rho=1$. Solid lines are
second order transitions with $z=2$; thick solid lines have $z=1$ and
appear to be in the KT universality class (see Eq.~(\ref{essential}));
the CDW-II to Mott-I transition is first order. Dashed lines connect
schematically the numerical results at the smallest $t/V$ to the exact
values at $t/V=0$.}
\label{phasediag}
\end{figure}

The phase diagram for a fixed value of $U$ in the $(\mu/t,\,t/V)$
plane is obtained by sweeping $\mu/t$ at constant $t/V$ as shown in
Fig.~\ref{rhomu}. This is shown in Fig.~\ref{phasediag} for
$U=10t$. The phases and the nature of the transitions between them are
explained in the caption. The details of the phase diagram above
$\rho=1$ have not been determined beyond establishing the natures of
the phases just above the CDW-II and Mott-I as indicated in the
figure. When CDW-II (Mott-I) is doped above or below $\rho=1$, the
system becomes supersolid (superfluid). We remark that in
figure~\ref{phasediag} ($U=10t$) the CDW-II and Mott-I lobes have a
very small contact region. This contact region increases as $U$ is
increased, {\it i.e.} the gap at the transition increases. On the
other hand, if $U$ is decreased, the two lobes separate and the direct
transition between CDW-II and Mott-I is lost: a SF phase intervenes.

Having shown that supersolid phases exist, we turn now to the
excitation spectrum, $\Omega(k)$ which, along with $\rho_s$,
characterizes the superfluid phase.  If $\Omega(k)\propto k$ for small
$k$, excitations are phonons and the superfluid is said to be stable
according to the Landau criterion: the Landau critical velocity,
$v^L_c$, is finite and is given by the slope of the line passing
through the origin and tangent to the roton (or, for $d=1$, the
soliton) minimum. To obtain $\Omega(k)$, one can calculate, via QMC,
the space and imaginary time separated density-density correlation
function which then gives the dynamic correlation function. With the
help of a numerical Laplace transform, for example using the Maximum
Entropy method, $\Omega(k)$ can be obtained~\cite{ggb4}. This
procedure is demanding numerically.  However, it was shown~\cite{ggb4}
that the rigorous Maxent procedure is not necessary since one obtains
a good approximation (upper bound) for $\Omega(k)$ by using the
$f$-sum rule leading to Feynman expression,
\begin{equation}
\Omega(k) = \frac{E_k}{L S(k)}.
\label{disp}
\end{equation}
where,
\begin{equation}
 E_k = \frac{-t}{L}\left ( {\rm cos}\,k-1\right ) \langle \Psi_0|
\sum_{i=1}^L \left ( a^{\dagger}_i a_{i+1} + a^{\dagger}_{i+1}
a_i\right)|\Psi_0 \rangle,
\end{equation}
and $|\Psi_0\rangle$ is the ground state of the system.

\begin{figure}
\psfig{file=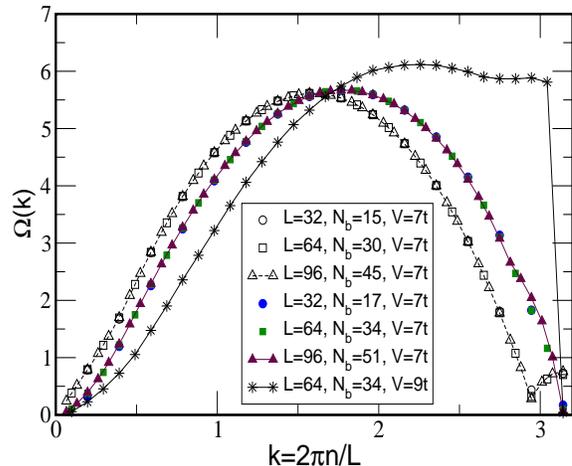,height=3.5in,width=3in,angle=-90}
\vskip-0.3in
\caption{The dispersion relation, $\Omega(k)$, vs the wave-vector,
  $k$, for several system sizes and boson numbers $N_b$. For all cases
  $U=10t$.}
\label{omega1d}
\end{figure}

Figure~\ref{omega1d} shows $\Omega(k)$ vs $k$ for several values of
the density and interaction strength. The first three cases listed in
the legend correspond to a fixed density $\rho=15/32$, just below half
filling in the SF phase. We note: (a) finite size effects are
negligible, (b) $\Omega(k)\propto k$ as $k\to 0$, indicating phonon
excitations and a stable SF, and (c) the soliton minimum is clearly
visible at $k=2.95$. As $\rho \to 1/2$ from below, $\Omega(k)$ remains
linear for small $k$ but the soliton minimum moves towards $k=\pi$ and
its energy approaches zero, pinching the $k$-axis, and signaling the
CDW instability. Therefore, the Landau critical velocity, $v^L_c\to 0$
although the SF remains stable, with a high phonon velocity, all the
way to the transition. The linearity of $\Omega(k)$ at the transition
shows that the SF-CDW transition has a dynamic critical exponent
$z=1$.

The next three cases listed in the legend, Fig.~\ref{omega1d},
correspond to $\rho=17/32$, just inside the SS phase. Again finite
size effects are under control. In addition, we see that $\Omega(k)$
{\it softens} as $k\to 0$ and is no longer linear and that this effect
increases with increasing $V$ (the last case in
Fig.~\ref{omega1d}). Fitting the form $\Omega(k)\propto k^z$ to the
small $k$ dispersion curves and sizes up to $L=96$ for $\rho=0.51$
gives $z\approx 1.9$. This suggests a dynamical critical exponent
$z=2$ for the CDW-SS transition. The same behavior is observed for
$\Omega(k)$ as the $\rho=1$ CDW-II phase (alternating doubly occupied
and vacant sites) is approached from the SS phase above or below full
filling, $\rho\to 1^{\pm}$.  Therefore, the quantum phase transitions
to the $\rho=1/2$ CDW are different depending on whether they are
approached from below ($\Omega(k\to 0)\propto k$ and soliton minimum
pinching the $k$-axis) or from above ($\Omega(k)$ softens to a
quadratic form).  However, the transitions to the CDW-II phase,
$\rho\to 1^{\pm}$, are in the same universality class with $z=2$.  We
also have performed simulations in two dimensions and verified that
the CDW-SS transition has $z=2$.

Therefore, $\Omega(k)$ yields for the SF-CDW transition $z=1$ while
$z=2$ for the SS-CDW transitions both at $\rho=1/2$ and $\rho=1$ (and
presumably also for $\rho=3/2$ etc). In addition, it was shown in
Ref.~\cite{fish,ggb5} that, when $V=0$, $\rho_s$ scales as $\rho_s\sim
|\rho_c-\rho|^{z-1}$ with $z=2$ where $\rho_c=1,2,..$ is the density
at which the MI phase is established. Our data show that for the
SS-CDW $\rho_s\sim |\rho_c-\rho|^{0.95}$ again giving $z\approx 2$
assuming the same scaling law. On the other hand, for the SF-CDW
transition $z=1$ yielding $\rho_s\sim |\rho_c-\rho|^0$ ($\rho_c=1/2$
here) which means that for this transition $\rho_s$ does {\it not}
vanish as a power law. In the absence of power law scaling, it is
natural to attempt a Kosterlitz-Thouless (KT) scaling form,
\begin{equation}
\rho_s = A {\rm e}^{-B/\sqrt{|\rho-\rho_c|}}
\label{essential}
\end{equation}
where $A$ and $B$ are nonuniversal constants. This is illustrated in
the inset of Fig.~\ref{rhomu} which shows $\rho_s$ vs $\rho$ for
$L=64,\, U=10t,\,V=8.5t$. The solid line is a two parameter fit with
Eq.~(\ref{essential}), the dashed line is a two parameter power law
fit to the same points. The goodness of this fit suggest that this
transition is in the KT universality class.

\begin{figure}
\psfig{file=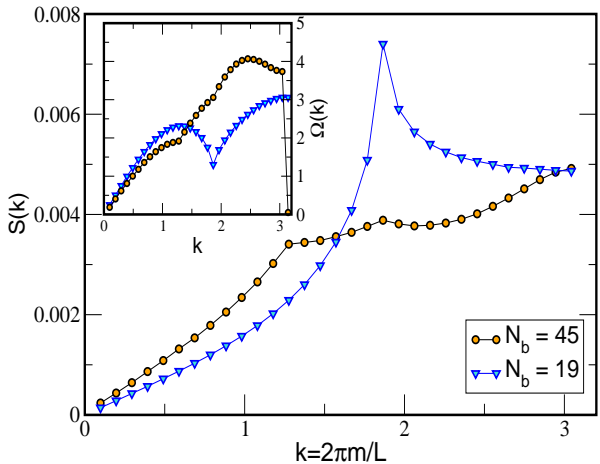,height=3.5in,width=3in,angle=-90}
\vskip-0.3in
\caption{Comparison of the structure factor, $S(k)$ for the same
  doping above and below half filling.  $L=64,\, \beta=20,\, U=10t,\,
  V=8t$. Inset: The corresponding dispersion, $\Omega(k)$.}
\label{Skcompare}
\end{figure}

The dispersion curves in Fig.~\ref{omega1d} show clearly that when the
system is doped below $\rho=1/2$, $\Omega(k)$ develops a soliton
dip. Figure~\ref{Skcompare} shows the origin of this behavior. For
$N_{b}=19$, $\Omega(k)$ shows a pronounced minimum, while for $N_b=45$
it shows two milder ones (inset Fig.~\ref{Skcompare}). These
excitation dips are caused by peaks in the structure factor which
point to the presence of quasi long range order with the corresponding
ordering vectors, ${\bf k}^*$. It is interesting to note that for
$N_b=45$ (and in general for $\rho>1/2$) the structure factor $S(k)$
has two peaks (Fig.~\ref{Skcompare}) indicating the simultaneous
presence of two types of quasi long range order. It is clear from the
figure that one of the peaks occurs at exactly the same ${\bf k}^*$ as
the soliton peak for the same doping {\it below} half filling.  This
peak for $\rho>1/2$ is caused by bosons occupying previously empty
sites thus aquiring near neighbors even though the large value of $V$
tries to suppress this. In other words, this is the vestige of the
approximate particle-hole symmetry present for $U=10t$ and $V\leq 6t$.
The peak at the lower $k$ comes from the quasi long range order
produced when doubly occupied sites are produced. Recall that at full
filling the ground state is a CDW with alternating doubly occupied and
empty sites.

We now return to the behavior of the compressibility,
$\kappa=\partial\rho/\partial\mu$, as $\rho\to 1/2^{\pm}$. It was
argued in~\cite{fish} for the SF-MI transition ($V=0,\, \rho=1$) that
$\kappa \sim |\mu-\mu_c|^{\nu(d-z)}$, where $\mu_c$ is the chemical
potential at the transition. For the $\rho\to 1/2^{+}$ transition,
$(d=1,z=2)$, our preliminary results indicate $\kappa \sim
|\mu-\mu_c|^{-0.6}$ which is consistent with $\nu=1/2$ assuming the
same scaling. For the $\rho\to 1/2^{-}$ transition $(d=1,z=1)$, this
scaling yields $\kappa \sim |\mu-\mu_c|^{0}$ which means that, although
$\kappa\to 0$ as $\rho \to 1/2^-$, it is not a power law.

In summary, we have presented QMC results showing that, contrary to
previous results~\cite{white} where $U>2V$, the extended
one-dimensional soft core bosonic Hubbard model does have SS phases
when doped above half filling when both $U$ and $V$ are large such
that the system favors multiple rather than near nighbor
occupancy. From the scaling of $\Omega(k)$, $\rho_s$ and $\kappa$, we
calculated the dynamic critical exponent, $z$, for all the
transitions. We showed different universality classes due to the
different behavior of $\Omega(k)$ depending on whether the CDW is
approached from below or above half filling and we highlighted
similarities and differences with the two dimensional case. Finally we
note that $1d$ SS phases have been predicted for commensurate mixtures
of two bosonic species, each near the hardcore limit, but where the
interaction between the species is not too large\cite{mathey}.

\noindent
\underbar{Acknowledgements} We thank M.~Troyer, A.~Sandvik and
G.~Carlin for helpful discussions.  
R.T.S. was supported by NSF-DMR-0312261.

\end{document}